\documentstyle[aps]{revtex}

\newcommand {\beq}{\begin{equation}}
\newcommand {\eeq}{\end{equation}}
\newcommand {\beqa}{\begin{eqnarray}}
\newcommand {\eeqa}{\end{eqnarray}}
\newcommand{\dal}{\fbox{\phantom{${\scriptstyle : }$}}}

\begin{document}
\title{The thermal and two-particle stress-energy must be
ill-defined on the 2-d Misner space chronology horizon }
\author{Claes R Cramer\cite{crc} and Bernard S. Kay\cite{bsk}}
\address{Department of Mathematics\\ University of
York,\\ Heslington\\ York YO1 5DD UK}
\maketitle
\begin{abstract}
We show that an analogue of the (four dimensional) image sum method can
be used to reproduce the results, due to Krasnikov, that for  the model
of a real massless scalar field on the initial globally hyperbolic
region $IGH$ of two-dimensional Misner space there exist  two-particle
and thermal Hadamard states (built on the conformal vacuum) such that
the (expectation  value of the renormalised) stress-energy tensor in
these states  vanishes on $IGH$. However, we shall prove that the
conclusions of  a general theorem by Kay, Radzikowski and Wald still
apply for these states.  That is, in any of these states, for any point
$b$ on the Cauchy horizon and any neighbourhood  $N$ of $b$, there exists
at least one pair of non-null related points  $(x,x')\in (N\cap
IGH)\times (N\cap IGH)$ such that (a suitably differentiated form of)
its two-point function is singular.  (We prove this by showing that
the two-point functions of these states share the same singularities as
the conformal vacuum on which they are built.) In other words, the
stress-energy tensor in any of these states is  necessarily ill-defined
{\it on} the Cauchy horizon.   
\end{abstract}

\hspace{1.4cm} PACS Number: 04.62.+v
\twocolumn
\section{INTRODUCTION}
The question whether or not an arbitrarily advanced civilization could 
possibly manufacture a time machine is still open. It has been argued 
that such a construction may be impossible, since the Cauchy/chronology 
horizon arising in any such attempt  would be semi-classically unstable,
see \cite{Thorne93} and references therein. It was thought that the
nature  of the semi-classical instability would be manifested by a
divergence in  (the expectation value of the renormalised) stress-energy
tensor as the  Cauchy horizon is approached. In fact, a simple example
of such a divergence  in the case of a real scalar field on Misner space
\cite{Hiscock82} was used by Hawking \cite{Hawking92} to argue that the
divergence may be universal, that is that a similar statement may hold
in any spacetime with a  (compactly generated) Cauchy horizon for any
physically acceptable initial  state.  We know now that this is not the
case: It has been shown by Sushkov \cite{Sushkov97} in the context of a
massless automorphic field on  four dimensional Misner space
\cite{Misfoot} that there exist  Hadamard states for which the
stress-energy tensor vanishes on the initial  globally hyperbolic
region.  Also, Krasnikov \cite{Krasnikov96} has given examples of
two-particle and thermal Hadamard\cite{massfoot} states 
(built on the conformal vacuum) for which the
stress-energy tensor  of a real massless scalar field on two-dimensional
Misner space vanishes  (or is bounded) on the initial globally
hyperbolic region.  (See also the earlier work of Sushkov
\cite{Sushkov95}.) Additionally,  Boulware \cite{Boulware92} in the
context of Gott space and Hiscock and  Tanaka \cite{Hiscock96} in the
context of Grant space\cite{Grantfoot} have  shown that for a
sufficiently massive real scalar field the stress-energy  tensor is
bounded on the initial globally hyperbolic region. Also,  Visser
\cite{Visser97} has argued that with a suitable wormhole  configuration
-- the Roman ring -- the stress-energy tensor can be made  arbitrarily
small all the way up to the Cauchy horizon.

On the other hand,  it has been shown rigorously by Kay, Radzikowski and
Wald (KRW) \cite{Kay97}  by using general theorems on `propagation of
singularities' \cite{Hormander85} that for the model consisting of a
real linear scalar field on a four  dimensional spacetime with compactly
generated Cauchy horizon  (or on a spacetime which arises as the product
of a Riemannian manifold of  dimension $4-d$ with a spacetime with
compactly generated Cauchy horizon with  dimension $d<4$) the
stress-energy tensor for any Hadamard state is  necessarily ill-defined
on the horizon.  (Technically, this ill-definedness is expressed
in terms of a singularity property of the state's two-point function.
In the present paper, it will be expressed in terms of a singularity
property of the state's point-split stress-energy tensor.)
By an obvious {\it reductio ad absurdum} argument this
result can be given the  interpretation that any spacetime with a
compactly generated Cauchy horizon  together with a real linear scalar
field cannot arise as a solution of  semi-classical gravity. 
(Suppose such a solution existed, then the stress-energy tensor would
be well-defined at every point of the spacetime!) Note that
if it were permissible to assume continuity  of the stress-energy tensor,
one could argue that when the stress-energy tensor  vanishes on the
initial globally hyperbolic region then it must be defined  and vanish
on the Cauchy horizon. The KRW-theorem shows, instead, that the 
stress-energy tensor is not defined there and of course one conclusion
from  this is that the assumption of continuity cannot be permissible.

We should  remark here that, even though the KRW-theorem by the above 
{\it reductio ad absurdum} argument can be given the status of a no-go 
theorem for semi-classically describable time machines, there still
remains the possibility that a full theory of quantum gravity might
still allow  time machines to exist in a (yet unknown) sense which is
not describable semi-classically. In fact, Visser \cite{Visser972} has
argued for an  invariant notion of {\it reliability horizon} as the
future boundary of the  region consisting of closed spacelike geodesics
with length  equal to or shorter than the  Planck length and argues that
we should not trust semi-classical gravity in  the region to the future
of the reliability horizon. Visser emphasizes that  the very existence
of models where the stress-energy can still be made small  up to the
Cauchy horizon of course prevents us from ruling out the possibility of
such  non-semi-classically describable time machines. However, it is not
at all clear what a time-machine would be once one loses the
notion of a classical spacetime, and even if there was such a notion,
what it would mean for a physical ``observer'' to pass through such
a time-machine.  Thus, in the absence of  a clearer
understanding of quantum gravity, a strong case can still be made for 
identifying  ``time-machine'' with ``semi-classically describable
time-machine'' whereupon the KRW-theorem may be interpreted as a no-go 
theorem. 

We remark that the KRW-theorem does not explicitly apply to  Sushkov's
\cite{Sushkov97} model consisting of a massless automorphic field on
four dimensional Misner space, but it has been proved  \cite{Cramer96}
in an elementary way that the conclusions of the  KRW-theorem apply to
Sushkov's model. Furthermore, the examples in  two-dimensions by
Krasnikov \cite{Krasnikov96} are of a slightly different nature from
the four dimensional examples, since the notion of Hadamard states on a
two-dimensional spacetime is  different from the corresponding notion in
four dimensions and because of infrared divergences, in the massless
theory, only derivatives of the field  make sense \cite{massfoot}.
Therefore, the KRW-theorem cannot be used  immediately to make a
statement on the nature of the stress-energy tensor on the Cauchy
horizon in a two-dimensional spacetime. However, we shall in this paper
prove, with a similar method to that used in  \cite{Cramer96}, that the
stress-energy tensor in the examples by Krasnikov must be ill-defined on
the Cauchy horizon.  

The organisation of this paper is as follows:  In section II we present
an independent derivation of the results of  Krasnikov
\cite{Krasnikov96} by pointing out that the two-point function of a
massless real scalar field in the conformal vacuum on two-dimensional
Misner space can be given an image sum form. In section III we  use the
fact that the conformal vacuum two-point function can be written as an
image sum to prove that the stress-energy tensor in the conformal vacuum
is ill-defined on the two-dimensional Misner space horizon. We then show
that, as a consequence of the result which tells us that the 
stress-energy tensor in the  conformal vacuum is ill-defined  on the
horizon, it must be similarly ill-defined for any (smooth) two-particle
or thermal Hadamard state built on the  conformal vacuum.

\section{THE CONFORMAL VACUUM IMAGE SUM}
Misner space (we mean here two-dimensional Misner space)
is a well known solution of Einstein's vacuum field equations 
with topology $\mathbf{R}\times S^1$, see e.g. \cite{Hawking73} and 
references therein. It consists of an initial globally hyperbolic 
region $IHG$ and a region of closed timelike curves $CTC$ separated by a 
compactly generated Cauchy horizon (closed nullgeodesic) $CH$. 
A mathematical description of Misner space can be given by taking the 
metric on the initial globally hyperbolic region $IGH$ to be 
\beq
ds^2=dt^2-t^2dx^2
\eeq
where $t\in (0,\infty)$ and $x$  is periodic with period $a$.
By going to double null coordinates (with an arbitrary
length scale $L>0$)
\beqa
u:=L\ln\frac{t}{L}-x\\
v:=L\ln\frac{t}{L}+x
\eeqa
with $u,v\in\mathbf{R}$ and where points are identified by 
$(u,v)\leftrightarrow (u-a,v+a)$, one finds that the globally hyperbolic 
region $IGH$ is conformal to the timelike cylinder, since
\beq
ds^2=e^{\frac{u+v}{L}}dudv\label{conmet}.
\eeq
It is convenient to introduce another set of coordinates 
$U,V\in (-\infty,0)$ defined by
\beqa
U:=-Le^{\frac{u}{L}}\\
V:=-Le^{\frac{v}{L}}.
\eeqa
The metric then takes the  form
\beq
ds^2=dUdV\label{minmet}
\eeq
and points are identified by a Lorentz boost with rapidity $a/L$, that is
$(U,V)\leftrightarrow (e^{-a/L}U,e^{a/L}V)$. The form of the metric and the
identification of points under a Lorentz boost allows one to conclude
that the initial globally hyperbolic region $IGH$ of Misner space can be
regared as the quotient of the region $\{(U,V):-\infty<U<0,-\infty<V<0\}$ 
of Minkowski space by a Lorentz boost. Furthermore, by extending the metric 
over the Cauchy horizon $CH$ and into the $CTC$ region  
$\{(U,V): -\infty<U<0,0<V<\infty\}$ of Misner space one easily concludes 
that Misner space is the quotient of the region 
$\{(U,V):-\infty <U<0,V\in \mathbf{R}\}$ of Minkowski space by a Lorentz 
boost.

We now turn to consider the model of a real massless scalar field $\phi$ 
on the initial globally hyperbolic region $IGH$ of Misner space satisfying 
the massless Klein-Gordon equation
\beq
\dal_g\phi=0.
\eeq
To quantise the system we use that the (conformal) positive frequency modes
\beq
u_k(u,v)= \left\{ 
\matrix{
(2a|k|)^{-1/2}e^{-ikv} &  k<0\cr
(2ak)^{-1/2}e^{iku} & k>0\cr}
\right.
\eeq
with the (complexified) Klein-Gordon inner product defines a one-particle 
Hilbert space $\mathcal{H}$ which in turn defines a Fock space 
$\mathcal{F}$ with the conformal vacuum $|C\rangle$, see e.g. \cite{Wald94} 
for details on one-particle Hilbert space structure. From the standard 
mode-sum expansion of the field $\phi$ in terms of the positive frequency 
modes and by calculations with distributions we find that the 
conformal vacuum two-point function $G_C$ in coordinates 
$x=(u,v)$ {\it formally} arises as the {\it image sum}
\beqa
G_C(x,x'):=\langle C |\phi(x)\phi(x')|C\rangle=\nonumber\\
-\frac{1}{4\pi}
\sum_{n=-\infty}^{\infty}\ln\{(u-u'+na)(v-v'-na)\}. 
\label{twop}
\eeqa
We remark that because of the problem with infrared divergences
(from another point of view, because the argument of the logarithm in
(\ref{twop}) should really be divided by an ill-defined length scale) in the
two-dimensional massless scalar field, only derivatives of the
two-point function $G_C$ make sense. Hence it is to be understood that
differentiation of the two-point function is performed before the summation
over images.  
The stress-energy tensor $\langle T_{ab}\rangle_C$ in the conformal vacuum 
is defined by
\beq
\langle T_{ab}\rangle_C:=\lim_{x,x'\rightarrow y} 
{\mathcal{T}}_{ab}(x,x')
\label{tab}
\eeq 
where the (renormalised) {\it point-split stress-energy tensor} 
${\mathcal{T}}_{ab}$ (here, thanks to flatness, primed and unprimed 
tensor indices can be identified) is defined by 
\beqa
{\mathcal{T}}_{ab}(x,x'):=\{ \partial_a\partial_{b'}-
\frac{1}{2}g_{ab}g^{cd'}\partial_{c}\partial_{d'}\}\nonumber\\
(G_C(x,x')-G_0(x,x'))\label{twotab}
\eeqa
and where
$$
G_0(x,x')=
-\frac{1}{4\pi}\ln\{L^2(e^{\frac{u}{L}}-e^{\frac{u'}{L}})
(e^{\frac{v}{L}}-e^{\frac{v'}{L}})\}
$$
is the usual Minkowski vacuum two-point function. By straightforward
calculations, one finds that the stress-energy tensor in the conformal vacuum 
in $(u,v)$ coordinates is 
\beq
\langle T_{ab}\rangle_C=
-(\frac{\pi}{12a^2}+\frac{1}{48L^2\pi})\hbox{diag}(1,1)
\eeq
or by going to $(U,V)$ coordinates
\beq
\langle T_{ab}\rangle_C=
-(\frac{\pi L^2}{12a^2}+\frac{1}{48\pi})
\hbox{diag}(U^{-2},V^{-2}).
\eeq
(in agreement with \cite{Krasnikov96}.)

The stress-energy tensor clearly diverges as one approaches the (future)
Cauchy horizon $(V=0,U=U_0<0)$. Nevertheless, following Krasnikov 
\cite{Krasnikov96} -- but continuing to use a different method -- we shall now  
show that there exist Hadamard states for which the stress-energy tensor 
vanishes on the initial globally hyperbolic region $IGH$. 
Consider a two-particle state in which the particles are in the 
momentum states $|n\rangle$ and $|-n\rangle$ then it is easy to show, 
using the mode sum 
expansion, that the two-point function $G_2$ in this state takes 
the form
\beqa
G_2(x,x')=G_C(x,x')
+2Re\{u_n(x)u^*_{n}(x')\}\nonumber\\
+2Re\{u_{-n}(x)u^*_{-n}(x')\}\label{twotwo}.
\eeqa 
Using (\ref{tab}) with $G_2$ replacing $G_C$  in (\ref{twotab}) we find 
that the stress-energy tensor $\langle T_{ab} \rangle_2$ in this two-particle 
state in $(u,v)$ coordinates is given by 
\beq
\langle T_{ab} \rangle_2=-
(\frac{\pi}{12a^2}+\frac{1}{48L^2\pi}
-\frac{2\pi n}{a^2})\hbox{diag}(1,1)
\eeq
and thus we have reproduced the result
of Krasnikov that (for any $L$) there exist choices of $a$ and $n$ such that 
the stress-energy 
tensor $\langle T_{ab} \rangle_2$ vanishes on the initial globally hyperbolic 
region $IGH$. Finally, consider a thermal state in which the 
thermal two-point function at inverse temperature $\beta$, $G_\beta$, 
is given by the image sum
\beqa
G_\beta(x,x')=
-\frac{1}{4\pi}
\sum_{m=-\infty}^{\infty}\sum_{n=-\infty}^{\infty} \{
\ln (u-u'+na-i\beta m)\nonumber\\+\ln (v-v'-na-i\beta m) \}
\label{thermtwo}.
\eeqa 
The thermal stress-energy $\langle T_{ab}\rangle_\beta$ is then obtained
by using (\ref{tab}) with $G_\beta$ replacing $G_C$ in (\ref{twotab}) 
which gives
\beq
\langle T_{ab}\rangle_\beta=-
(\frac{\pi}{12a^2}+\frac{1}{48L^2\pi}-\frac{\pi}{2a^2})
\sum_{m=1}^\infty\sinh^{-2} \frac{\pi m\beta}{a}\hbox{diag}(1,1)
\eeq
and thus we have reproduced the result of Krasnikov
that (for any $L$) there exist choices of $a$ and $\beta$ such 
that the stress-energy tensor $\langle T_{ab}\rangle_\beta$ vanishes on 
the initial globally hyperbolic region $IGH$.
\section{THE STRESS-ENERGY TENSOR IS ILL-DEFINED {\it ON} THE HORIZON}
We shall now present an elementary proof of a result (the theorem below) 
which tells us that in the case of a massless real 
scalar field on two-dimensional Misner space the stress-energy tensor 
$\langle T_{ab}\rangle_C$ in the conformal vacuum is necessarily 
ill-defined on the Cauchy Horizon.  Further, we shall show that,
as a consequence of this result, the stress-energy 
tensor $\langle T_{ab}\rangle_2$ in the two-particle state and the thermal 
stress-energy tensor $\langle T_{ab}\rangle_\beta$ are also ill-defined on 
the Cauchy horizon.  In particular, and this is the main point of the present
paper, even for the states discussed in the previous section, for which
the stress-energy tensor vanishes in the initial globally hyperbolic 
region $IGH$, it is still ill-defined {\it on} the Cauchy horizon
(similarly to the results in \cite{Kay97,Cramer96}).\\\\
{\bf Theorem}. Let $b$ be any point on the Cauchy horizon $CH$, and
$N$ be any neighbourhood of $b$. Then there exists at least one-pair
of non-null related points $(x,x')\in (N\cap IGH)\times (N\cap IGH)$, 
where $IGH$ is the initial globally hyperbolic region, such that the 
(renormalised) point-split stress-energy tensor ${\mathcal{T}}_{ab}(x,x')$ 
in the conformal vacuum $|C\rangle$ is singular.\\
{\bf Proof}. The (renormalised) point-split stress-energy tensor 
${\mathcal{T}}_{ab}(x,x')$
is given by formulae (\ref{twotab})
\beqa
{\mathcal{T}}_{ab}(x,x')= \hspace{2cm}\nonumber\\
\{\partial_a\partial_{b'}-
\frac{1}{2}g_{ab}g^{cd'}\partial_{c}\partial_{d'}\}
(G_C(x,x')-G_0(x,x'))\nonumber
\eeqa
where the two-point functions $G_C$ and $G_0$ in coordinates $(U,V)$ are 
given by
\beqa
G_C(x,x')=
-\frac{1}{4\pi}
\sum_{n=-\infty}^{\infty}\{\ln(L\ln(\frac{-U}{L})-L\ln(\frac{-U'}{L})+na)
\nonumber\\
+\ln(L\ln(\frac{-V}{L})-L\ln(\frac{-V'}{L})-na)\}\nonumber
\eeqa
and
$$
G_0(x,x')=-\frac{1}{4\pi}\ln\{(U-U')(V-V')\}.
$$
{}From the form of the metric (\ref{minmet}) and the two-point functions 
$G_C$ and $G_0$ it follows that the only non-vanishing components of the 
point-split stress-energy tensor are
\beqa
{\mathcal{T}}_{UU}(x,x')=-\hspace{3cm}\nonumber\\\frac{1}{4\pi}
(\sum_{n}
\frac{L^2}{UU'(L\ln(\frac{-U}{L})-L\ln(\frac{-U'}{L})+na)^2}-\frac{1}{(U-U')^2}
)\nonumber 
\eeqa
and
\beqa
{\mathcal{T}}_{VV}(x,x')=-\hspace{3cm}\nonumber\\
\frac{1}{4\pi}
(\sum_{n}
\frac{L^2}{VV'(L\ln(\frac{-V}{L})-L\ln(\frac{-V'}{L})-na)^2}-
\frac{1}{(V-V')^2}).\nonumber 
\eeqa
Clearly $N\cap IGH$ will contain an open rectangle 
$R=\{(U,V):U_0<U<U_1,V_0<V<0\}$. Thus it will be sufficient to show that 
there exists a pair of points in the open rectangle $R$ having the 
properties stated in the theorem. Let $x=(U,V)\in R$. Then it is easy to 
see that there exists a point $\hat x=(\hat U,\hat V)\in R$ spacelike 
separated from $x$ such that every point 
$x_\delta=(\hat U, \delta)$, $\hat V <\delta<0$ along the line connecting 
$\hat x$ to the point $(\hat U, 0)$ on the Cauchy horizon is also spacelike 
separated to $x$. We now show that we can choose $\delta$ so that the 
point-split stress-energy tensor ${\mathcal{T}}_{ab}(x,x')$ is singular. 
{}First we note that for any $\delta$ which takes the form $\delta=Ve^{-ma/L}$ 
the $m$th term in the sum in the expression above for the component 
${\mathcal{T}}_{VV}(x,x_\delta)$ will diverge and by taking $m$ to be 
sufficiently large and positive we can arrange for $\delta$ to lie in the 
interval $(\hat V,0)$. Taking $x'=x_\delta$ we have thus exhibited a pair
of points in $(N\cap IGH)\times (N\cap IGH)$ such that the mth term in the 
sum of ${\mathcal{T}}_{VV'}(x,x')$ is singular. On the other hand, it is easy 
to see that the remaining terms of the sum are uniformly convergent in some 
neighbourhood of $(x,x')$. Thus we conclude that the point-split
stress-energy tensor ${\mathcal{T}}_{ab}(x,x')$ is singular.\\\\
{\bf Corollary 1}. The statement of the theorem is true for the two-particle 
point-split stress-energy tensor
\beqa
{\mathcal{T}}_{ab}(x,x')=\hspace{2cm}\nonumber\\
\{\partial_a\partial_{b'}-
\frac{1}{2}g_{ab}g^{cd'}\partial_{c}\partial_{d'}\}
(G_2(x,x')-G_0(x,x')).\nonumber
\eeqa
{\bf Proof}. The proof is obvious from the facts that the two-point function 
$G_2(x,x')$ (\ref{twotwo}) is given by
\beqa
G_2(x,x')=G_C(x,x')\hspace{1cm}\nonumber\\
+2Re\{u_n(x)u^*_{n}(x')\}
+2Re\{u_{-n}(x)u^*_{-n}(x')\}\nonumber
\eeqa
and that the positive frequency modes $u_{\pm n}$ are smooth on $IGH$.

We remark that this corollary can in a trival way be extended to any
many particle state built (by acting with creation operators
smeared with smooth wave-functions) on the conformal vacuum, since the 
two-point function in such a state will consist of the conformal vacuum 
two-point function plus a smooth two-point function.\\\\ 
{\bf Corollary 2}. The statement of the theorem is true for the thermal 
point-split stress-energy tensor 
\beqa
{\mathcal{T}}_{ab}(x,x')=
\hspace{2cm}\nonumber\\
\{\partial_a\partial_{b'}-
\frac{1}{2}g_{ab}g^{cd'}\partial_{c}\partial_{d'}\}
(G_\beta(x,x')-G_0(x,x')).\nonumber
\eeqa
{\bf Proof}.
The proof is immediate from the theorem and the observation that the 
thermal two-point function $G_\beta(x,x')$ (\ref{thermtwo}) can be 
written as 
\beqa
G_\beta(x,x')=G_C(x,x')-\hspace{1.5cm}\nonumber\\
\frac{1}{4\pi}\sum_{m\not =0}
\sum_{n}\{
\ln(L\ln(\frac{-U}{L})-L\ln(\frac{-U}{L})'+na-i\beta m)+\nonumber\\
\ln(L\ln(\frac{-V}{L})-L\ln(\frac{-V'}{L})-na-i\beta m)\}.
\nonumber
\eeqa
and since the differentiated form of the (double) sum is uniformly 
convergent (to a {\it smooth} limit).

It is clearly an immediate consequence of the above theorem and
corollaries that the limit in equation (\ref{tab}), and in the
corresponding equations for $\langle T_{ab}\rangle_2$ and $\langle
T_{ab}\rangle_\beta$, is ill-defined, thus we have shown what we
set out to show.

{}Finally we remark, that it is clear 
that a similar theorem holds in the context of massless automorphic fields on 
two-dimensional Misner space, since it can be shown \cite{Cramer97} that 
the two-point function of a massless automorphic field takes the same
form as the conformal two-point function $G_C$  with the exception that an 
{\it automorphic} factor of $\exp (2\pi i \alpha n)$ is inserted 
into the image sum.  (That the case of four-dimensional Misner space is similar
was shown in \cite{Cramer96}.)
\\\\
\acknowledgements          
CRC is grateful to the WW Smith Bequest of the University of York for
partial financial support.


\begin{references}
\bibitem{Misfoot}
The Cauchy horizon of four dimensional Misner space is strictly speaking not 
compactly generated but arises as the product of a Euclidian plane with 
two-dimensional Misner space.
\bibitem{Grantfoot}
We remark that neither Grant nor Gott space has a compactly generated
Cauchy horizon or arises as the product of a $d-4$ dimensional Riemannian 
manifold with a spacetime of dimension $d<4$ with compactly generated Cauchy 
horizon.
\end{references}
\end{document}